\documentclass[a4paper,11pt]{article}

\usepackage{amsmath,amsfonts,amssymb,caption,graphicx}
\usepackage{ushort}
\usepackage{makeidx}
\usepackage{float}
\usepackage{accents}
\usepackage{color}
\usepackage{framed}
\usepackage{versions}
\usepackage{xcolor}
\usepackage{emptypage}
\usepackage{wrapfig}
\usepackage[T1]{fontenc}
\usepackage[utf8]{inputenc}
\usepackage{titlesec}
\usepackage{fancyhdr}
\usepackage{extramarks}
\usepackage[bookmarks]{hyperref}
\usepackage{hyperref}
\usepackage{bbm}
\usepackage[baseline]{euflag}

\hypersetup{
    colorlinks=true,   	
    linkcolor=red,      
    citecolor = [rgb]{0 0.7 0},   	
    filecolor=magenta, 	
    urlcolor=blue
}

\newcommand{\RL}{{\mathbb R}}

\newcommand{\Phatn}{\mbox{$\hat{P}_n$}}

\def\ba{\begin{align}}
\def\ea{\end{align}}
\def\ban{\begin{align*}}
\def\ean{\end{align*}}

\def\be{\begin{eqnarray}}
\def\ee{\end{eqnarray}}
\def\ben{\begin{eqnarray*}}
\def\een{\end{eqnarray*}}

\def\bqq{\begin{equation}}
\def\eqq{\end{equation}}
\def\bqqn{\begin{equation*}}
\def\eqqn{\end{equation*}}

\def\sq{$\Box$}

\def\qed{\ifmmode\sq\else{\unskip\nobreak\hfil
\penalty50\hskip1em\null\nobreak\hfil\sq
\parfillskip=0pt\finalhyphendemerits=0\endgraf}\fi\par\medbreak}

\newsavebox{\junk}
\savebox{\junk}[1.6mm]{\hbox{$|\!|\!|$}}

\def\til={{\widetilde =}}

\def\clA{{\cal A}}

\def\clF{{\cal F}}
\def\clG{{\cal G}}

\def\clM{{\cal M}}

\def\clP{{\cal P}}

 \def\eq#1/{(\ref{#1})}

\newtheorem{theorem}{Theorem}[section]
\newtheorem{corollary}[theorem]{Corollary}
\newtheorem{proposition}[theorem]{Proposition}
\newtheorem{lemma}[theorem]{Lemma}

\def\eq#1/{(\ref{e:#1})}

\def\bdes{\begin{description}}
\def\edes{\end{description}}

\def\notes#1{}

\definecolor{mag}{rgb}{0.7,0,0.3}
\definecolor{dgreen}{rgb}{0.1,0.5,0.1}
\definecolor{dred}{rgb}{.8,0,0}
\definecolor{gray}{rgb}{.8,.8,.8}
\definecolor{brown}{rgb}{0.6451,0.3706,0.1745}

\begin{document}
 
\title{\vspace{-1.1cm}%
A Third Information-Theoretic Approach\\ to Finite de Finetti Theorems}

\author
{
	Mario Berta
   \thanks{Department of Computing, Imperial College, 
	London SW7 2AZ, U.K.,
	and Institute for Quantum Information, RWTH Aachen University, 
	Aachen, Germany.
                Email: \texttt{\href{berta@physik.rwth-aachen.de}%
			{berta@physik.rwth-aachen.de}}.
	M.B. acknowledges funding by the European Research Council 
	(ERC Grant Agreement No.~948139).
	}
\and
	Lampros Gavalakis
    \thanks{Univ Gustave Eiffel, Univ Paris Est Creteil, CNRS, LAMA UMR8050 F-77447 Marne-la-Vall{\'e}e, France.
                Email: \texttt{\href{mailto:lampros.gavalakis@univ-eiffel.fr}%
			{lampros.gavalakis@univ-eiffel.fr}}.
	L.G. was supported in part by the European Union's Horizon
	2020 research and innovation program under the Marie 
	Sklodowska-Curie grant agreement No 101034255 \euflag \ and by the B{\'e}zout Labex, funded by ANR, reference ANR-10-LABX-58.
        }
\and
        Ioannis Kontoyiannis 
    \thanks{Statistical Laboratory, DPMMS,
	University of Cambridge,
	Centre for Mathematical Sciences,
        Wilberforce Road,
	Cambridge CB3 0WB, U.K.
                Email: \texttt{\href{mailto:yiannis@maths.cam.ac.uk}%
			{yiannis@maths.cam.ac.uk}}.
	I.K.\ was supported in part by the Hellenic Foundation for Research 
	and Innovation (H.F.R.I.) under the ``First Call for H.F.R.I. Research 
	Projects to support Faculty members and Researchers and the 
	procurement of high-cost research equipment grant,'' project 
	number 1034.
        }
}

\date{\today}

\maketitle

\begin{abstract}
A new finite form of de Finetti's representation theorem
is established using elementary information-theoretic tools.
The distribution of the first $k$ random variables in an
exchangeable vector of $n\geq k$ random variables
is close to a mixture of product distributions. 
Closeness is measured in terms of the relative entropy 
and an explicit bound is provided. This bound is tighter
than those obtained via earlier information-theoretic
proofs, and its utility extends to random variables
taking values in general spaces.
The core argument employed has its
origins in the quantum information-theoretic
literature.
\end{abstract}


\newpage

\section{Introduction: de Finetti's theorem}

A vector of $n$ random variables
$(X_1,X_2,\ldots,X_n)$ is {\em exchangeable} 
if it has the same distribution as 
$(X_{\pi(1)},X_{\pi(2)},\ldots,X_{\pi(n)})$ 
for every permutation $\pi$ on $\{1,2,\ldots,n\}$. 
Similarly, we say that a process $\{X_k\;;\;k\geq 1\}$ 
is exchangeable if $(X_1,X_2,\ldots,X_n)$ is
exchangeable for all $n$.

De Finetti's celebrated representation 
theorem~\cite{definetti:31,definetti:37} 
characterises all binary exchangeable processes.

\begin{theorem}[de Finetti's representation theorem]
\label{definettiintro} 
A binary process $\{X_k\;;\;k\geq 1\}$ is exchangeable 
if and only if its distribution can be 
uniquely expressed
as a mixture of independent and identically
distributed binary processes.
\end{theorem}

The importance of de Finetti's theorem (and 
its numerous extensions) for the foundations of Bayesian 
statistics is illustrated, e.g., by the observation 
that it justifies the subjective approach of treating 
any infinite binary exchangeable sequence 
as a sequence of {\em independent} coin tosses conditional
on the value of the probability of Heads, which is itself
distributed according to some prior distribution --
namely, the mixing measure in Theorem~\ref{definettiintro};
see~\cite{diaconis:77,bayarri:04}
and the references therein for extensive discussions.
 
In terms of its practical applicability in statistics
and elsewhere, it is natural to ask whether the same 
representation applies to finite-length exchangeable 
random vectors. As it turns out, the condition that the
entire process be exchangeable is necessary. For example,
considering a pair of binary random variables $(X_1,X_2)$
with $\Pr(X_1 = 1, X_2 = 0) = \Pr(X_1=0,X_2 = 1) = 1/2$,
shows that $(X_1,X_2)$ are exchangeable but their joint
distribution cannot be expressed as a mixture of 
independent and identically distributed (i.i.d.)
binary pairs~\cite{diaconis:77}.

Nevertheless, the distribution of the first $k$ random 
variables in an exchangeable vector of length $n$ admits 
an {\em approximate} de~Finetti-style representation if $k$ is 
small compared to $n$.
Quantitative versions of this statement have been
established by Diaconis~\cite{diaconis:77} for binary 
random variables
and Diaconis and Freedman~\cite{diaconis-freedman:80b} for 
random variables with values 
in abstract spaces, with sharp error bounds for 
the total variation distance. We refer to such results 
as `finite de Finetti' theorems. 
Diaconis' proof in~\cite{diaconis:77}
is based on a geometric interpretation of the set of 
exchangeable measures as a convex subset of the probability 
simplex. The Diaconis and Freedman proof
in~\cite{diaconis-freedman:80b} is based on the 
approximation of the hypergeometric by the binomial 
distribution. The general version of 
Theorem~\ref{definettiintro} due to Hewitt and 
Savage~\cite{hewitt-savage:55} for random variables
with values in compact Hausdorff spaces
was also recovered in \cite{diaconis-freedman:80b}. 
Interestingly, one needs additional assumptions 
in order to deduce the infinite version from the 
-- seemingly stronger -- finite one. 

\subsection{Information-theoretic approaches}

Since the early 1950s, there has been a long line
of works establishing core probabilistic results
via information-theoretic ideas and techniques;
parts of this rich history are outlined
in~\cite{csiszar:97,barron:97,johnson:book,gavalakis-LNM:23}.
In this spirit, two different {\em information-theoretic} proofs 
of finite  de Finetti theorems were developed recently,
one for binary alphabets~\cite{gavalakis:21} and 
one for finite alphabets~\cite{gavalakis-LNM:23}. 
These results are expressed in terms of the relative entropy,
which is a stronger notion of `distance' than total variation,
but the resulting total variation bounds are generally
suboptimal. 
We briefly recall these results to highlight their
differences with the present development; a short 
review of these two approaches 
may be found in \cite{gk:ITW}. 

For any $k$, write $X_1^k$ for a vector
of random variables $(X_1,X_2,\ldots,X_k)$
and similarly write $x_1^k$ for a specific
realisation $(x_1,x_2,\ldots,x_k)$ of $X_1^k$.
When the $X_i$ all take values in a 
finite alphabet, we denote
their joint probability mass function
by $P_{X_1^k}$.
If $\mu$ is a probability measure on the
simplex $\clP$ of probability mass functions
$Q$ on $A$, we define the mixture:
$$M_{k,\mu}(x_1^k):=\int_{\cal P} Q^k(x_1^k)\,d\mu(Q),
\quad x_1^k\in A^k.$$

The main result of the first information-theoretic
proof~\cite{gavalakis:21} is that,
if $X_1^n$ is a binary exchangeable random vector,
then there is a probability measure 
$\mu$ on the space of Bernoulli distributions
such that, for every $k \leq n$,
\bqq
D(P_{X_1^k}\|M_{k,\mu}) \leq \frac{5k^2\log n}{n-k},
\label{eq:first}
\eqq
where
$D(P\|Q)=\sum_{x\in B}P(x)\log[P(x)/Q(x)]$
denotes the relative entropy between two probability
mass functions on the same finite alphabet~$B$,
and $\log$ denotes the natural logarithm throughout.
Writing $\|P-Q\|:=\sup_B |P(B)-Q(B)|$ 
for the total variation distance
between $P$ and~$Q$,
Pinsker's inequality~\cite{csiszar:67,kullback:67} 
states that
$\|P-Q\|^2\leq \frac{1}{2}D(P\|Q)$.
Thus,~(\ref{eq:first}) yields,
\begin{equation} 
\label{eq:l1bound}
\|P_{X_1^k} - M_{k,\mu} \| 
\leq k\Bigl(\frac{5\log{n}}{2(n-k)}\Bigr)^{\frac{1}{2}}.
\end{equation}
This bound is suboptimal in that,
as shown in~\cite{diaconis-freedman:80b},
the correct rate with respect to the total variation 
distance in~(\ref{eq:l1bound}) is $O(k/n)$.
The proof in~\cite{gavalakis:21} 
is based on an estimate of the dependence 
between the random variables $X_1^k$
conditional on the proportion of 1s it contains.

In the second information-theoretic approach to 
finite de Finetti theorems~\cite{gavalakis-LNM:23},
a different finite-$n$ bound was obtained for the relative 
entropy in~(\ref{eq:first}), leading to an estimate
of the form:
\bqq
D(P_{X_1^k}\| M_{k,\mu})=
O\left(\Big(\frac{k}{\sqrt{n}}\Big)^{1/2}\log{\frac{n}{k}}\right).
\label{eq:second}
\eqq
Although~(\ref{eq:second}) is generally weaker than~(\ref{eq:first}),
it holds for random vectors $X_1^n$ with values in arbitrary finite
alphabets.
The proof of the finite-$n$ bound leading to~(\ref{eq:second}) 
in~\cite{gavalakis-LNM:23}
was based on a connection between 
the Gibbs conditioning principle from statistical mechanics,
the information-theoretic `method of types',
and de Finetti's theorem.

Finally, we note that a simple argument 
was recently used~\cite{gavalakis-olly:pre}
to show that, for finite-valued
exchangeable random vectors, it is possible
to obtain a bound of the form:
\bqq
D(P_{X_1^k}\| M_{k,\mu})=O\Big(\Big(\frac{k}{n}\Big)^2\Big),
\label{eq:stam}
\eqq
where the coefficient is proportional to the alphabet size, $ |A|$.
In view of the
$O(k/n)$ lower bound~\cite{diaconis-freedman:80b}
for the total variation distance mentioned 
above, Pinsker's inequality implies that the
rate,
in terms of $k$ and $n$,
achieved in~(\ref{eq:stam})
is in fact optimal 
for the relative entropy.
Unfortunately, the proof of~(\ref{eq:stam})
in~\cite{gavalakis-olly:pre} is {\em not}
information-theoretic, as it is based on
an earlier bound by Stam~\cite{stam:78}
for the relative entropy 
between the distributions of sampling 
with and without replacement,
which was established by direct probabilistic
and combinatorial arguments. For the sake of
completeness, we state and prove~(\ref{eq:stam})
in Section~\ref{s:lower}.

\subsection{A different information-theoretic approach}

The main goal of this note is to present and explore
yet another information-theoretic approach
to developing finite de Finetti theorems, based only
on elementary properties of the mutual information. 
To put it in context, we first remark that
de Finetti-style theorems have attracted increasing 
attention in the literature on quantum information 
theory in recent years. 
For example, starting from related ideas 
in~\cite{brandao:11}, the third part 
of~\cite[Theorem~1]{brandao:13CMP} 
may be viewed as a classical de Finetti theorem 
with additional constraints, and a 
classical de Finetti result may also 
be obtained from the subsequent 
work~\cite[Theorem~2.4]{berta:22}.
Additional relevant results can be found 
in~\cite{brandao:13bCMP,borderi:phd,li:15} and 
the references therein.

One of our results, given as
Corollary~\ref{corcountable} in the following
section, states that, if $X_1^n$ is an exchangeable 
vector of random variables taking values in finite
alphabet $A$, then:
$$D(P_{X_1^k}\|M_{k,\mu}) \leq \frac{k(k-1)}{2(n-k-1)}\log |A|.$$
This bound, which is stronger 
than those obtained in~\cite{gavalakis:21} and 
in~\cite{gavalakis-LNM:23},
will be derived
as a consequence of our main 
result, Theorem~\ref{relativeentropybound}.

The main idea of the proof of the general
bound in Theorem~\ref{relativeentropybound}
has its origins in quantum information theory.
It is based on an argument that was implicitly 
used in the proof of the quantum
information-theoretic result in
\cite[Theorem~1]{brandao:13CMP}.
The fact that this argument leads to interesting
classical de Finetti-style
results appears to have been noted by several
authors, and it is even mentioned as an exercise
in the recent text~\cite{polyanskiy-wu:pre}.

Our main purpose here is to explore and 
adapt this argument in order to show that it
can be employed to obtain
finite-$n$ bounds for de Finetti's
theorem in terms of relative entropy,
which are both stronger and more general
than those obtained via the earlier 
information-theoretic 
proofs~\cite{gavalakis:21,gavalakis-LNM:23}.

Compared to the Diaconis and Freedman~\cite{diaconis-freedman:80b}
variational distance bound $k(k-1)/(2n)$, 
our information-theoretic corollary is
weaker in terms of the dependence on $n$ and $k$, and additionally features
the logarithm of the alphabet size. However, our bound holds in the stronger
relative entropy distance and it is then an interesting 
question if the logarithm
of the alphabet size is needed for that distance measure. Moreover,
following~\cite{berta:22,brandao:13CMP}, we note that our bound also holds
for de Finetti theorems with linear constraints. In the same way that
Diaconis and Freedman's result~\cite{diaconis-freedman:80} gives a polynomial
time approximation scheme for the minimization of polynomials of fixed
degree over the simplex ~\cite{deklerk:06,deklerk:15}, the versions with
linear constraints from~\cite{berta:22} -- that feature the logarithm
of the alphabet size -- give rise to quasi-polynomial time approximation
schemes for the minimization of polynomials of fixed degree over the simplex
intersected with affine constraints~\cite{berta:22}. Additionally, 
for this quasi-polynomial
time approximation scheme there also exists a matching
hardness-of-approximation result~\cite{aaronson:14}. As such, 
we conclude that at least
under certain complexity-theoretic assumptions, any finite de Finetti proof
strategy that can incorporate linear constraints will likely need 
to feature the logarithm
of the alphabet size in the approximation error.

\section{Finite de Finetti theorems}

\subsection{A general bound under exchangeability}

We begin with some preliminary definitions and
assumptions.

Let $(A, \clA)$ be a measurable space. 
Write $\clM_1(A)$ for the space of probability measures 
on $(A, \clA)$.
For $P,Q \in \clM_1(A)$, the {\em relative entropy}
between $P$ and $Q$ is,
\begin{equation*} 
D(P\|Q) =
\begin{cases} 
    \int{\log{\frac{dP}{dQ}\;dP}},& \text{if } P \ll Q,\\
    +\infty,              & \text{otherwise},
\end{cases}
\end{equation*}
where $dP/dQ$ denotes the 
Radon-Nikod{\'y}m derivative of $P$ with respect to $Q$.
For a pair of random variables $(X,Y)$ with values
in the measurable spaces $(A_1,\clA_1)$ and $(A_2,\clA_2)$, 
respectively, and with joint law
$P_{X,Y} \in \clM_1(A_1 \times A_2)$,
the {\em mutual information between $X$ and $Y$} is,
\begin{equation*}
I(X;Y) = D(P_{X,Y}\|P_X \times P_Y),
\end{equation*}
where $P_X\in\clM_1(A_1)$ and 
$P_Y\in\clM_1(A_2)$ denote the first 
and second marginal of $P_{XY}$, respectively.
In order to consider mixture measures 
on given measurable space $(A,\clA)$,
we first need to equip $\clM_1(A)$ with its own
$\sigma$-algebra. For every $F\in\clA$, define
the map $\pi_F:\clM_1(A)\to[0,1]$ by,
$$\pi_F(P)=P(F),\quad P\in\clM_1(A),$$
and let $\clF$ denote the smallest $\sigma$-algebra
of subsets of $\clM_1(A)$
that makes the maps $\{\pi_F\;;\;F\in\clA\}$
measurable. Now consider a specific
measure $\mu\in\clM_1(\clM_1(A))$
and a given $k\geq 1$. For any $F\in\clA^k$
the map $Q\mapsto Q^k(F)$ is $\clF$-measurable,
so we can define the mixture
$M_{k,\mu}\in\clM_1(A)$ as,
$$M_{k,\mu}(F):=\int_{\clM_1(A)} Q^k(F)\,d\mu(Q),
\quad F\in\clA^k.$$
Throughout, we assume that $(A,\clA)$ is 
standard a Borel space, namely, that there exists a complete 
separable metric space whose Borel $\sigma$-algebra 
is isomorphic to $\mathcal{A}$. This ensures that 
regular conditional probabilities exist \cite{parthasarathy:book}.
Although in many cases, especially in potential
applications of our results,
this assumption may be unnecessary, it ensures that, in the general
case, we will not be committing any measure-theoretic
{\em faux pas}.

We can now state our main result.

\begin{theorem}[Exchangeability and information]
\label{relativeentropybound}
Suppose $X_1^n$ is an exchangeable vector
of random variables $X_i$ with values in
a standard Borel space  $(A,\clA)$,
and with joint law $P_{X_1^n}$.
For every $1\leq k\leq n-1$ there exists
a probability measure $\mu=\mu_{k,n}$ on 
$\clM_1(A)$, such that:
\begin{equation} 
\label{mutualinfosbound}
D\big(P_{X_1^k} \big\|M_{k,\mu}\big) 
\leq  \frac{1}{n-k+1} \sum_{i=1}^k{I(X_1^{i-1};X_k^n).}
\end{equation}
\end{theorem}

For the proof we will need the following two lemmas.
The first one is a simple identity for the
relative entropy, stated without proof.
The second one provides a lower bound
on the mutual informations
between $X_1^{i}$ and $X_k^n$, for $k>i$,
which will be a key step
in the proof of Theorem~\ref{relativeentropybound}.

\begin{lemma}
\label{lem:Drepresent}
For any random vector $Z_1^k$ of length $k$, 
the relative entropy between the joint law $P_{Z_1^k}$ 
of $Z_1^k$ and the product $\prod_{i=1}^kP_{Z_i}$
of its marginals can be expressed as:
\begin{equation*}
D\Big(P_{Z_1^k}\Big\|\prod_{i=1}^kP_{Z_i}\Big)
=\sum_{i=1}^k I(Z_1^{i-1};Z_i).
\label{eq:DIexp1}
\end{equation*}
\end{lemma}

Recall that
conditional mutual information between $X$ and $Y$ 
given $Z$ is defined as,
\begin{equation*}
I(X;Y|Z) 
= \int{
D\Bigl(P_{X,Y|Z}(\cdot|z)\Big\|P_{X|Z}(\cdot|z)\times P_{Y|Z}(\cdot|z)\Bigr) 
\,dP_Z(z)},
\end{equation*}
where $P_{X,Y|Z}(\cdot|z), P_{X|Z}(\cdot|z)$, $P_{Y|Z}(\cdot|z)$,
denote the conditional laws of $(X,Y)$, $X$, $Y$, given $Z$,
respectively, whenever these exist.

\begin{lemma}
\label{lem:IHbound}
Suppose $X_1^n$ are 
as in Theorem~\ref{relativeentropybound}.
Then, for each $1\leq i \leq k\leq n-1$:
\begin{equation*}
\sum_{m=k}^n{I(X_1^{i-1};X_i|X_{k+1}^m)} \leq I(X_1^{i-1};X_k^n).
\end{equation*}
\end{lemma}

\noindent
{\sc Proof. }
Let $1\leq i \leq k\leq n-1$ be fixed.
For any $k\leq m\leq n$, by exchangeability we have,
\begin{equation*} 
I(X_1^{i-1};X_i|X_{k+1}^{m}) = I(X_1^{i-1};X_{m}|X_{k}^{m-1}),
\end{equation*}
with the obvious convention that $X_{k+1}^k$ is just a constant.
Summing
over all $k\leq m\leq n$ and using the chain
rule for mutual information yields,
\begin{equation*} 
\sum_{m = k}^{n}{I(X_1^{i-1};X_{i}|X_{k+1}^m)} = 
\sum_{m = k}^{n}{I(X_1^{i-1};X_{m}|X_k^{m-1})} =
I(X_1^{i-1};X_k^n),
\end{equation*}
as claimed.
\qed

\noindent
{\sc Proof of Theorem~\ref{relativeentropybound}. }
Adding the identities of Lemma~\ref{lem:IHbound}
over all $1\leq i\leq k$
and dividing by $(n-k+1)$,
$$
\frac{1}{n-k+1}\sum_{m=k}^n{\sum_{i=1}^k{I(X_1^{i-1};X_i|X_{k+1}^m)}} 
\leq \frac{1}{n-k+1} \sum_{i=1}^k{I(X_1^{i-1};X_k^n).} 
$$
Since the average over $m$ satisfies this upper bound, there is an $m^* 
\in \{k,k+1,\ldots,n\}$ such that,
\bqq
\sum_{i=1}^k{I(X_1^{i-1};X_i|X_{k+1}^{m^*})}
\leq \frac{1}{n-k+1} \sum_{i=1}^k{I(X_1^{i-1};X_k^n).}
\label{eq:average}
\eqq
Let $Q^*=Q^*_{k,n}\in\clM_1(A^{m^*-k})$ denote the joint law of $X_{k+1}^{m^*}$.
Taking $\mu=\mu_{k,n}$ to be the law of the regular conditional 
probability $P_{X_1|X_{k+1}^{m^*}}(\cdot|X_{k+1}^{m^*})$,
the mixture $M_{k,\mu}$ can be expressed:
$$M_{k,\mu}(F)
=\int_{A^{m^*-k}}P_{X_1|X_{k+1}^{m^*}}^k(F|x_{k+1}^{m^*}) 
\,dQ^*(x_{k+1}^{m^*}),
\quad F \in \clA^k.
$$
Using the joint convexity of relative entropy,
\begin{align*}
D\big(P_{X_1^k} \big\| M_{k,\mu}\big) 
&=
	D\Big(
	\int_{A^{m^*-k}}P_{X_1^k|X_{k+1}^{m^*}}(\cdot|x_{k+1}^{m^*}) 
	\,dQ^*(x_{k+1}^{m^*})
	\Big\|
	\int_{A^{m^*-k}}P_{X_1|X_{k+1}^{m^*}}^k(\cdot|x_{k+1}^{m^*}) 
	\,dQ^*(x_{k+1}^{m^*})
	\Big) \\
&\leq 
	\int
	D\Big(
	P_{X_1^k|X_{k+1}^{m^*}}(\cdot|x_{k+1}^{m^*})
	\Big\|
	P_{X_1|X_{k+1}^{m^*}}^k(\cdot|x_{k+1}^{m^*})
	\Big)\,dQ^*(x_{k+1}^{m^*})\\
&=
	\int
	D\Big(
	P_{X_1^k|X_{k+1}^{m^*}}(\cdot|x_{k+1}^{m^*})
	\Big\|
	\prod_{i=1}^kP_{X_i|X_{k+1}^{m^*}}(\cdot|x_{k+1}^{m^*})\Big)
	\,dQ^*(x_{k+1}^{m^*}),
\end{align*}
where the last step follows from the fact that,
by exchangeability,
$P_{X_i|X_{k+1}^{m^*}}$ is the same for all $i\leq k$.
Finally, applying Lemma~\ref{lem:Drepresent}
to the integrand,
yields,
$$
D\big(P_{X_1^k} \big\| M_{k,\mu}\big) 
\leq 
	\sum_{i=1}^k{I(X_1^{i-1};X_i|X_{k+1}^{m^*})},$$
and the claimed bound follows from this combined with~(\ref{eq:average}).
\qed

\subsection{Explicit de Finetti-style theorems}

For random variables $X_1^k$ on a discrete (finite or countably infinite)
alphabet~$A$, the mutual information always satisfies~\cite{cover:book2},
$$I(X_1^k;Y)=H(X_1^k)-H(X_1^k|Y)\leq H(X_1^k)\leq k\log |A|,$$
where $H(X)=-\sum_{x\in B}P(x)\log P(x)$ denotes the entropy
of a random variable with probability mass function
$P$ on a discrete alphabet $B$. Therefore, Theorem~\ref{relativeentropybound}
immediately yields:

\begin{corollary}[Finite de Finetti theorem for discrete random variables]
\label{corcountable}
\,
Suppose $X_1^n$ is an exchangeable vector of
random variables $X_i$ taking values in a discrete alphabet~$A$.
For every $1\leq k\leq n-1$ there exists a probability
measure $\mu=\mu_{k,n}$ on 
$\clM_1(A)$, such that:
$$
D\big(P_{X_1^k} \big\|M_{k,\mu}\big) 
\leq \frac{k(k-1)}{2(n-k+1)}H(X_1) \leq \frac{k(k-1)}{2(n-k+1)}\log |A|.
$$
\end{corollary} 

Compared with the earlier information-theoretic
results~(\ref{eq:first}) and~(\ref{eq:second}),
the bound in Corollary~\ref{corcountable} is both more general 
and stronger. Moreover, it can be used
to recover the classical infinite version of de Finetti's theorem 
for compact spaces, under some conditions.

\begin{corollary}[Classical de Finetti theorem for compact spaces]
\label{cor:compactG} \,
Let $G$ be a compact metrisable space equipped with its 
Baire $\sigma$-algebra $\clG$. Suppose the process
$\{X_k\;;\;k\geq 1\}$ is exchangeable and the
$X_k$ take values in $G$ and are $\clG$-measurable.
If for every $k$ we have $I(X_1^{k-1};X_k^n)=o(n)$ as $n\to\infty$, 
then there is a probability measure $\mu$ on $\clM_1(G)$ such that:
$$
P_{X_1^k} = M_{k,\mu},\quad \mbox{for every}\; k\geq 1.
$$
\end{corollary} 

Note that Corollary~\ref{cor:compactG}
applies to all finite-alphabet exchangeable processes:
Any finite set $G=\{a_0,a_1,\ldots,a_{m-1}\}$ 
is compact with respect to the 
metric $d(a_i,a_j)=|i-j|$ (mod~$m$).
And since
$I(X_1^{k-1};X_k^n)\leq H(X_1^{k-1})\leq
(k-1)\log m$,
the condition 
$I(X_1^{k-1};X_k^n)=o(n)$ is always
satisfied.

\noindent
{\sc Proof. }
Choose and fix $k\geq 1$.
Since $G$ is compact and metrisable, it is also complete 
and separable and thus standard Borel.
And since $\{X_k\}$ is exchangeable,
we can apply
Theorem~\ref{relativeentropybound} 
in combination with 
Pinsker's inequality to obtain that,
for any $n\geq k$,
\bqq
\|P_{X_1^k} -M_{k,\mu_{k,n}} \|
\leq  \left[\frac{1}{2(n-k+1)} \sum_{i=1}^kI(X_1^{i-1};X_k^n)\right]^{1/2}.
\label{eq:DiFr}
\eqq
The condition $I(X_1^{k-1};X_k^n)=o(n)$ implies that
the right-hand side vanishes as $n \to \infty$
and we are almost done, except that the mixing measure
$\mu_{k,n}$ depends on both $n$ and $k$.

Since $G$ is compact and metrisable, it is also Hausdorff.
Following~\cite{diaconis-freedman:80b}, we note
that 
$\clM_1(\clM_1(A))$ is 
compact in the weak* topology,
so that there is a $\mu_k\in\clM_1(\clM_1(A))$ and
a subsequence $\{n_j\}$ 
increasing to infinity such that
$\mu_{k,n_j}\to \mu_k$ as $j\to\infty$.
Then, since the map $\mu\mapsto M_{k,\mu}$
is weak*-continuous,
we have that
$M_{k,\mu_{k,n_j}}\to M_{k,\mu_k}$
in the weak* topology
as $j\to\infty$, and, in view of~(\ref{eq:DiFr}),
we also have that 
$M_{k,\mu_{k,n_j}}\to
P_{X_1^k}$ in total variation.
Therefore, we must have that,
$$P_{X_1^k}=M_{k,\mu_k}.$$
Since $k$ was arbitrary, by the fact that 
the marginals of the process $\{X_k\}$ are
necessarily consistent we must also have
that
$P_{X_1^k}=M_{k,\mu_n}$,
for each $k$ and all $n\geq k$.
Using weak*-compactness again, we 
can find a subsequence $\{\mu_{k_\ell}\}$
that converges to some $\mu\in\clM_1(\clM_1(A))$
in the weak* topology as $\ell\to\infty$.
By the consistency of the marginals
and using again the weak*-continuity of $M_{k,\mu}$
in $\mu$, we then have that for each $k$,
as $\ell\to\infty$,
$P_{X_1^k}=M_{k,\mu_{k_\ell}}\to M_{k,\mu}$
in the weak* topology, completing the proof.
\qed

\subsection{Examples}

Except for finite-valued processes
(noted after Corollary~\ref{cor:compactG})
the condition $I(X_1^{k-1};X_k^n) = o(n)$ 
appears rather technical and is generally not easy
to verify. The following example 
describes a class of real-valued exchangeable 
processes that satisfy the condition 
$I(X_1^{k-1};X_k^n) = o(n)$.
Although these are not covered 
by Corollary~\ref{cor:compactG}
or the classical infinite de Finetti 
theorems (since $\RL$ is not compact), 
our Theorem~\ref{relativeentropybound} 
does provide useful bounds.

\medskip

\noindent
{\bf Example~1. }
Consider a finite collection of densities $\{f_1,f_2,\ldots,f_m\}$
on $\RL$, such that the differential entropy
$h(f_i)=-\int f_i\log f_i$ exists and is finite
for each $i$. Let $\theta$ be an arbitrary random variable
with values in $\{1,2,\ldots,m\}$.
Define an exchangeable real-valued process $\{X_k\;;\;k\geq1\}$
as follows. Conditional on $\theta=i$, let
$\{X_k\}$ be i.i.d.\ with each $X_k\sim f_i$.
Then, since conditioning reduces
the differential entropy~\cite{cover:book2},
$$
I(X_1^{k-1};X_k^n) 
= h(X_1^{k-1}) - h(X_1^{k-1}|X_k^n) 
\leq h(X_1^{k-1}) - h(X_1^{k-1}|X_k^n,\theta),
$$
and since $X_1^{k-1}$ is conditionally independent of $X_k^n$
given $\theta$,
$$I(X_1^{k-1};X_k^n) 
= h(X_1^{k-1}) - h(X_1^{k-1}|\theta) 
\leq (k-1)\bigl[h(X_1) - \min_{1\leq i\leq m}{h(f_i)}\bigr].
$$ 
Therefore, 
in this case
Theorem~\ref{relativeentropybound} 
provides a useful bound of $O(k/n)$ for the relative
entropy.

\medskip

On the other hand, the following is 
an example of an exchangeable process 
$\{X_n\}$ on a compact space, 
for which the condition
$I(X_1^{k-1};X_k^n) = o(n)$
fails.

\medskip

\noindent
{\bf Example 2. }
Define a process $\{X_k\;;\;k\geq 1\}$ as follows.
Let $\theta\sim U[0,1]$ and, conditional on $\theta=x$,
let the $X_k$ be i.i.d.\ with each $X_k$ taking the 
values $x/2$ and $x$ with probability 1/2 each.
Then $\{X_k\}$ is a mixture of i.i.d.\ processes and
hence exchangeable. But
it is easy to see that the joint law
$P_{X_1^n}$ is never absolutely continuous
with respect to $P_{X_1^{k-1}}\times P_{X_k^n}$.
Therefore, not only do we not have
$I(X_1^{k-1};X_k^n) = o(n)$,
but $I(X_1^{k-1};X_k^n)$ is equal to $+\infty$
and the result 
of Theorem~\ref{relativeentropybound}
is trivial. One possible explanation for this
is that,
in some cases, the mixing measure $\mu$ obtained
in the proof 
of Theorem~\ref{relativeentropybound}
is not the ``right'' one.

\subsection{A non-information-theoretic bound}
\label{s:lower}

Finally, we prove the upper bound~(\ref{eq:stam})
in the Introduction.

\smallskip

\begin{proposition}[Optimal de Finetti upper 
bound~\cite{gavalakis-olly:pre}]
If $X_1^n$ is an exchangeable vector of random variables 
$X_i$ with values in a finite alphabet $A$, then,
for every
$1\leq k\leq n - 1$, there exists a probability measure 
$\mu=\mu_{k,n}$ on $\clM_1(A)$, s.t.:
$$D\big(P_{X_1^k}\|M_{k,\mu}\big)
\leq\frac{1}{2} (|A| -1)\frac{k(k-1)}{(n-1)(n-k+1)}.
$$
\end{proposition}

\noindent
{\em Proof. }
Write $A=\{a_1,a_2,\ldots,a_m\}$
with $|A|=m$,
let $\Phatn$ denote the (random) empirical 
probability mass function (p.m.f.) induced 
by $X_1^n$ on $A$,
and let $\mu$ denote the distribution
of $\Phatn$ on $\clP$.
Write $\clP_n\subset\clP$ for the collection
of all possible p.m.f.s that can arise
as empirical distributions of strings
$x_1^n\in A^n$.
A key observation here is that,
for an exchangeable $X_1^n$,
conditional on $\Phatn=Q$
for some $Q\in\clP_n$,
the distribution of $X_1^n$ is uniform
over all $x_1^n$ with the same
empirical frequencies as $Q$,
namely, containing $nQ(a_j)$ appearances
of $a_j$, $j=1,2,\ldots,m$.
Therefore, conditional on $\Phatn=Q$,
the distribution of $X_1^k$
is the distribution of $k$ draws 
without replacement from 
an urn with $n Q(a_j)$ balls 
labelled $j=1,2,\ldots,m$.
Let $P^{({\rm nr})}_Q$ denote this distribution,
for each $Q\in\clP_n$, so that,
$$P_{X_1^k}=\sum_{P\in\clP_n}\mu(Q) P^{({\rm nr})}_Q.$$
We will compare this with the mixture of i.i.d.s,
$$M_{k,\mu}=\int Q^{k}d\mu(Q)
=\sum_{Q\in\clP_n}\mu(Q)Q^k
=\sum_{Q\in\clP_n}\mu(Q)P_Q^{({\rm r})},$$
where
$P_Q^{({\rm r})}=Q^k$ is the 
distribution of $k$ draws 
draws {\em with} replacement from 
an urn with $=n Q(a_j)$ balls 
labelled $j$ for each $j=1,2,\ldots,m$.
Now we recall the following bound due to
Stam~\cite{stam:78},
\bqq
D\big( P^{({\rm nr})}_Q\big\|P_Q^{({\rm r})}\big)
\leq\frac{(m-1)k(k-1)}{2(n-1)(n-k+1)}.
\label{eq:stamell}
\eqq
By the joint convexity 
of the relative entropy in its two arguments
we have,
\begin{align}
D(P_{X_1^k} \|M_{k,\mu})
&=
D\Big(\sum_{P\in\clP_n}\mu(Q) P^{({\rm nr})}_Q
\Big\|
\sum_{Q\in\clP_n}\mu(Q)P_Q^{({\rm r})}\Big)
\nonumber\\
&\leq
\sum_{P\in\clP_n}\mu(Q) 
D\big(P^{({\rm nr})}_Q\|
P_Q^{({\rm r})}\big),
\label{eq:prestam}
\end{align}
and the result follows upon
combining~(\ref{eq:stamell}) with~(\ref{eq:prestam}).
\qed

\def\cprime{$'$}

\end{document}